\documentclass[twoside,twocolumn]{article}

\usepackage{blindtext} 
\usepackage{array}
\newcolumntype{L}{>{\centering\arraybackslash}m{3cm}}

\usepackage[sc]{mathpazo} 
\usepackage[T1]{fontenc} 
\usepackage{microtype} 

\usepackage[english]{babel} 

\usepackage[hmarginratio=1:1,top=32mm, left=20mm, bottom=25mm, columnsep=20pt]{geometry} 
\usepackage[hang, small,labelfont=bf,up,textfont=it,up]{caption} 
\usepackage{booktabs} 

\usepackage{lettrine} 

\usepackage{enumitem} 
\setlist[itemize]{noitemsep} 

\usepackage{abstract} 

\usepackage{titlesec} 
\renewcommand\thesection{\Roman{section}} 
\renewcommand\thesubsection{\roman{subsection}} 
\titleformat{\section}[block]{\large\scshape\centering}{\thesection.}{1em}{} 
\titleformat{\subsection}[block]{\large}{\thesubsection.}{1em}{} 

\usepackage{fancyhdr} 
\usepackage{titling} 

\usepackage{hyperref} 

\usepackage{amssymb}
\usepackage{amsmath}
\usepackage{graphicx}

\pretitle{\begin{center}\huge\bfseries} 
\posttitle{\end{center}} 
\title{Differential Privacy and Natural Language Processing to Generate Contextually Similar Decoy Messages in Honey Encryption Scheme} 
\author{%
\textsc{Kunjal Panchal}\\
\normalsize University of Massachusetts, Amherst \\ 
\normalsize \href{mailto:kpanchal@umass.edu}{kpanchal@umass.edu} 
}
\date{\today} 
\renewcommand{\maketitlehookd}{%
\begin{abstract}
\noindent 
Honey Encryption is an approach to encrypt the messages using low min-entropy keys, such as weak passwords, OTPs, PINs, credit card numbers. The ciphertext is produces, when decrypted with any number of incorrect keys, produces plausible-looking but bogus plaintext called ``honey messages''. But the current techniques used in producing the decoy plaintexts do not model human language entirely. A gibberish, random assortment of words is not enough to fool an attacker; that will not be acceptable and convincing, whether or not the attacker knows some information of the genuine source.\\\\
In this paper, I focus on the plaintexts which are some non-numeric informative messages. In order to fool the attacker into believing that the decoy message can actually be from a certain source, we need to capture the empirical and contextual properties of the language. That is, there should be no linguistic difference between real and fake message, without revealing the structure of the real message. I employ natural language processing and generalized differential privacy to solve this problem. Mainly I focus on machine learning methods like keyword extraction, context classification, bags-of-words, word embeddings, transformers for text processing to model privacy for text documents. Then I prove the security of this approach with $\epsilon$-differential privacy.
\end{abstract}
}

\begin{document}

\maketitle


\section{Introduction}



\lettrine[nindent=0em,lines=3]{A}s the computer hardware (GPUs and mining rigs) got ``cheaper'' and widely available; difficulty of password cracking got more dependent on number of characters in a character encoding scheme, the lentgh of the password and hashing scheme. The possibilities of combinations of passwords are: 
$combination = \#char^{|password|}$. Table \ref{tab:mdl} shows comparison between two hashing schemes: MD5 and SHA1, which were analyzed in HashCat \cite{Cyberpunk:2020} benchmark to correlate different types of hashes on a machine with two Sapphire R580(8GB) cards.\\
\begin{table*}[t]
\begin{tabular}{l||ll}
                                                           & MD5                                                     & SHA                                                     \\ \hline\hline
Message Digest Length                                      & 128 bits                                                & 160 bits                                                \\ \hline
\multicolumn{1}{m{4cm}||}{Attacks required to find original message} & \multicolumn{1}{m{5cm}}{2\textasciicircum{}128 bit operations required to break} & \multicolumn{1}{m{5cm}}{2\textasciicircum{}160 bit operations required to break} \\ \hline
\multicolumn{1}{m{4cm}||}{Attacks to try and find two messages producing the same MD} & 2\textasciicircum{}64 bit operations                    & 2\textasciicircum{}80 bit operations                    \\ \hline
Speed                                                      & 64 iterations (faster)                                  & 80 iterations                                          
\end{tabular}
\caption{Comparison of MD5 and SHA1\cite{Cyberpunk:2018}}
\label{tab:mdl}
\end{table*}

In theory, with a 20726 MH/s machine, we can crack a 8 all alphabets lowercase characters MD5 password in 2 hours [$26^8/(20726\times10^6)\approx9500 seconds$]. With SHA1, it would be approximately 7.5 hours [$26^8/(7328.7\times10^6)\approx27410 seconds$]. \cite{Cyberpunk:2018}\\\\
Even the stupendous guessing rates do not yet make it feasible for an attacker to use a dumb brute-force technique of trying all possible combinations of characters to discover a password. Instead, password crackers rely on the fact that some people choose easily guessable passwords \cite{Stallings:2011}. Some users, when permitted to choose their own password, pick one that is absurdly short.\\\\
The results of one study at Purdue University observed that among approximately 7000 users, almost 3\% of the passwords are of length of 8 characters. Password length is only a part of a much larger issue. Many users, when allowed to pick their own password, pick a password that is easily guessable, like their own name, their street name, or a common dictionary word. This makes the job of password cracking straightforward. The cracker simply has to test the password file against lists of likely passwords. Because many people use guessable passwords, such a strategy should succeed on virtually all systems. There exists password attacks like ``dictionary attack'' where we just check each word of a certain high frequency wordlists; ``hybrid/rules attack'' where we augment those wordlists by adding characters before or after the high frequency words; ``markov attack'' where machine learning techniques are used to assemble possible guesses; ``mask attacks'' where the focus of the attack is on the structure of a password's character composition; and finally, ``brute force attack'' where we try every possible combination of characters. If the adversary knows the length of a password to be significantly short (PINs are normally of 4 digits, OTP lengths are publicly known for each system, generally 6, minimum length of passwords are mostly 8 or 12), then brute forcing the key space would not take very long \cite{Dunning:2015}\\\\
Bottom line is, users need to improve their password selection strategy through user education about network security, use of computer generated passwords, reactive or proactive password checking by the system. But some the password crackers are getting more sophisticated and smart too; and where the length of the password is limited, we face a roadblock.\\\\
Honey Encryption [HE] \cite{Ristenpart:2014} provides security beyond the brute-force bound. It is a scheme where adversary is not able to succeed in recovering the message even after trying every possible combination of password. But it is difficult to apply where the context of the message is known; e.g., adversary is trying to brute force into a military website, or the adversary knows that on a certain time, Alice and Bob are going to exchange messages about a certain topic, the messages will be confidential, but not the subject of the messages - then for a false password attempt, if our HE scheme returns a grammatically incoherent, gibberish message about ``toys'', adversary might not be fooled and will keep brute-forcing it until he finds a contextually similar or relevant message. That way adversary can get hold of the correct password combination.\\\\
To tackle this problem, Juels in \cite{Chatterjee:2015} discussed how strength of HE depends on fake but semantically and contextually realistic natural language decoy message. So here, I propose a method to generate realistic and deceitful messages on false password attempts. First, we need to classify the message into a category like ``politics'', ``sports'', ``military'', ``traveling'' and so on; the owner of message is free to manually classify the message into one of the options. This classification is needed to glean the semantic context of the message. To classify the messages, we need to extract keywords from the message. Later, with WordNets, we can uniformly randomly pick words with similar embeddings which will work as our new keywords for the fake text synthesis process. These will be mostly nouns and adjectives, because they can specify more context than verbs, which are almost always common and can be used with any kind of subject.\\\\
This will give us a bag-of-word representation. We can either generate deepfake text (a portmanteau of ``deep learning'' and ``fake'') from a two-layer neural network called Word2Vec or for much better realism, we can employ a transformer like GPT-2 \cite{Radford:2019} or Grover \cite{Zellers:2019}.\\\\
Next, I prove the security of the approach with ideas inspired by $\epsilon$-differential privacy, a metric-based extension of differential privacy \cite{Fernandes:2019}. I will define semantic similarity in terms of embedding score and Earth Mover's distance used for machine learning for classification. ``$E_\epsilon$-privacy'' is related to the Earth Mover's metric and is used for text obfuscation problems faced in the concept of differential privacy.


\section{Prior Work}
There's a comprehensive review of the HE scheme \cite{Abiodun:2018} where the most discussed issue with HE scheme is supposed to be designing a human and machine convincing decoy text without revealing the structure of the actual message. Basically, our goal is to design a fake text, which if juxtaposed with the actual text, the adversary (human or machine) will have negligible probability of differentiating between. The paper on how to synthesize decoy text \cite{Omolara:2018, Jantan:2019} discussed its use in HE scheme where the content of messages are textual in nature. The approach is semantically sound and coherent, but it does not address the issue of adversary having prescience about the context of the message. It only works for the situations where there is zero foreknowledge of the messages.\\\\
The paper on cracking resistant password vaults \cite{Chatterjee:2015} with Natural Language Encoder [NLE] deals with ``decoy passwords'' in a sense that if a master password to some vault is given false (as a consequence of a brute-force attack), the passwords inside the vault will be generated by an NLE, where a password having a natural language word will be encoded into some other password, without revealing its structure. But as the passwords are always context-free (only few words at most) and they have a certain probability distribution; this approach is hard to transfer into our ``decoy text'' problem. There's an extensive study of probabilistic password models \cite{Ma:2014} is given which supports the NLE approach of \cite{Chatterjee:2015}. This is useful when studying the nature of distribution-transforming encoders [DTE] used in original HE scheme \cite{Jaeger:2016}.\\\\
An alternative approach \cite{Ghassami:2016} to HE scheme addresses the loss in security due to non-uniformity of the decoy messages. It maps DTE's randomness to a larger set of messages which are nearly uniform. But its security is not always ``nearly achieving the information-theoretic converse on the probability of the success of the adversary'', it doesn't establish the precise performance of the new scheme. They lack the converse bounds needed for the comparison.\\\\
An article on universality of structured passwords \cite{Dunning:2015} shows how top 13 unique mask structures make up 50\% of the passwords from the sample size of over 34 million publicly exposed passwords which included famous password dumps. It gives a deep insight into the statistics like having an uppercase requirement, over 90\% of the time it is put as the first character. This kind of statistical password cracking bolsters the need of brute-force attack safe encryption schemes even more.\\\\
GPT-2 \cite{Radford:2019} is a large transformer-based language model with 1.5 billion parameters, trained on a dataset of 8 million web pages. When it generates text, it uses that knowledge to make the sentences more realistic. Therein lies the power of GPT-2: the text it generates actually contains real world facts and figures. Grover is a modified version of GPT-2 where the next predicted word is not only based on the previous words in the text, but also based on other pieces of metadata text \cite{Zellers:2019}.\\\\
Finally, \cite{Fernandes:2019} discusses the security through differential privacy in textual data.


\section{Goal}
First I define few concepts which we are using in our approach:\\\\
\textbf{DTE: } Distribution-Transforming Encoder is an encoding scheme where the randomized algorithm \textit{encode} takes an input message $M\in\mathcal{M}$ and outputs a seed value from a set of seed space $\mathcal{S}$. The deterministic algorithm \textit{decode} takes a seed value $S\in\mathcal{S}$ and outputs a message $M\in\mathcal{M}$. IF $Pr[decode(encode(M))=M]=1$, then the DTE scheme is correct.\\\\
Let $\mathcal{A}$ be an adversary attempting to distinguish between two DTE games where encoding game $G1$ randomly takes $M\in\mathcal{M}$ and a seed based on $M$; it produces a bit $b$, where $b = 1$ if $\mathcal{A}$ maps the ciphertext to correct seed, message pair. And the decoding game $G0$ randomly takes $S\in\mathcal{S}$ and a message decoded from that seed. It produces a bit $b$, where $b = 1$ if $\mathcal{A}$ maps the ciphertext to correct seed, message pair.\\\\
$Adv^{dte}_{DTE, p_m}(\mathcal{A}) = |Pr[G1^{\mathcal{A}}_{DTE, p_m}\implies 1]-Pr[G0^{\mathcal{A}}_{DTE}\implies 1]|$\\\\
\textbf{Framework of HE schemes: } For encryption, we get a password and a plaintext message; we generate a seed encoding from the message, and a uniformly random binary string. The ciphertext is an ``or''ed value between the seed and the keyed hash of the random string where the key is our password. A pair of the random string and ciphertext is returned.\\\\
For decryption, we get a password and the random string and ciphertext pair. We compute a seed by ``or''ing the ciphertext with the keyed hash of the random string we get as the input, where the key is the password. We decode the seed to a message. That message is the output, the decrypted text.\\\\
\textbf{NLTK: } Natural Language Toolkit provides corpora of different categories and lexical resources like WordNet, Word2Vec which are useful in finding contextually similar words.\\\\ 
\textbf{Text Classification: } This is where I assign a category to our message, according to its content. NLTK's brown corpus \cite{Brown:1991} is a million-word corpus in English. It has 46 subcategories, few of them being ``Political'', ``Sports'', ``Society'', ``Financial'', ``Government Documents'', ``College Catalog'' and such. Our goal is to classify our message into one of these categories, because our keyword extraction will depend on the classified category.\\\\ 
\textbf{Keyword Extraction: } This is a task where from a text, the terms that best describe the context of it are automatically identified. Extraction is a process where we choose the terms which are explicitly mentioned in the original text.\\\\
In information retrieval, Term Frequency-Inverse Document Frequency [TF-IDF] score shows how important a word is in a given categorical corpus of a text. It increases as the number of times a term shows up in a text increases, and that is offset by the number of documents in the classified corpus. A survey \cite{Breitinger:2015} showed that 83\% of text-based recommender systems in digital libraries use TF-IDF.\\\\
\textbf{Word2vec: }Word embedding is a mathematical embedding where words are mapped into a continuous vector space - $Vec: \mathcal{S}\rightarrow\mathbb{R}^n$, given a pseudorandom distribution on $\mathbb{R}^n$. The similarity of two words will be in $\mathcal{S}\times\mathcal{S}\rightarrow\mathbb{R}$ as $dist_{Vec}(w_1, w_2) := 1 -  dist(Vec(w_1), Vec(w_2))$. It aims to quantify and categorize semantic similarities between natural language terms based on their distribution in the classfied corpus. As John Firth said, ``a word is characterized by the company it keeps''. This mapping is mostly done with neural networks, a team at Google in 2013 created a toolkit named Word2vec, which can train vector space models faster than previous approaches using n-gram models or unsupervised learning. It takes a corpus as input, and produces embedding score of each terms.\\\\
Point of this model is that it will produce a similar embedding score for the terms used in similar context. This is what we need for our contextually similar decoy message.\\\\
\textbf{GPT-2: } It is a generative model of a language. It is a unsupervised, statistical model of how English is used on the Internet. When we feed it a few words, it starts predicting what the next word is most likely to be. And it takes all the history of text generated so far into account too. So it doesn't deviate a lot from the context. A transformer module encodes the meaning of each word based entirely on the words around it in the current sentence. Using sentence context is what makes it work so well. For more detail about the working and usage, reader can refer \cite{Geitgey:2019} \\\\
\textbf{Grover: } It was released at University of Washington. It is a modified version of GPT-2. It has better mechanisms for a recurrent neural network with ``attention'' as it also depends on metadata of the terms we want to synthesize text on.\\\\
\textbf{Differential Privacy: } If $M$ and $M'$ are classified to be similar in context, then depending on a private parameter $\epsilon$ the outputs determined by a contextually similar words finding mechanism $K$, will be $K(M)$ and $K(M')$ and they are also similar to each other, regardless of the authorship \cite{Fernandes:2019}. Possible results of $K$ are determined by the distribution of the Laplace probability density function quantified by the scores of word embeddings generated by Word2vec and a similarity metric called Earth Mover's distance.\\\\
\textbf{Earth Mover's Distance: }If $d_\mathcal{S}$ is a metric over $\mathcal{S}$. The Earth Mover's Distance with respect to $d_\mathcal{S}$, $E_{d_\mathcal{S}}$ is a  solution to a linear optimization problem: $E_{d_\mathcal{S}} := \min\sum_{x_i\in X} \sum_{y_j\in Y}d_\mathcal{S}(x_i, y_j)F_{ij}$ where $\sum_{i=1}^kF_{ij} = b_j/|Y|$ and $\sum_{j=1}^lF_{ij} = a_i/|X|$.\\\\
\textbf{Earth Mover's Privacy: }If $\mathcal{S}$ is a set, $d_\mathcal{S}$ is a pseudo-metric on $\mathcal{S}$ and $E_{d_\mathcal{S}}$ is a Earth Mover's Distance on $\mathcal{S}$ relative to $d_\mathcal{S}$. Given $\epsilon\geq0$, a contextually similar words finding mechanism $K: \mathcal{S}\rightarrow\mathbb{D}(\mathcal{S})$ satisfies $\epsilon E_{d_\mathcal{S}}$-privacy if and only is for any $M$, $M' \in \mathcal{S}$ and $Z\subseteq\mathcal{S}$ there is $K(M)(Z)\leq e^{\epsilon E_{d_\mathcal{S}}(M,M')}K(M')(Z)$.\\\\
It says that if two messages are calculated to be very close, that $\epsilon E_{d_\mathcal{S}}(M, M')$ is close to 0, then $e^{\epsilon E_{d_\mathcal{S}}(M,M')}$ wil be approx 1. And thus, the outputs are almost identical. And if two messages are more distinguishable, the more their exponents will differ. For the proof, refer to \cite{Fernandes:2019}.\\\\
Now I can move ahead with putting all these pieces together to form an efficient decoy message generation scheme. Let's think of two scenarios:\\\\
1. Adversary knows or expects a certain subject for the chosen ciphertext attack.\\\\
2. Adversary does not have any knowledge of the subject of the ciphertext.\\\\
For the second case, even if the adversary does not and should not know the semantic context of the actual message $M$, just sending a random message $M'$ everytime will  diminish the indistinguishableness of the scheme. If the words of two different categories are outputted for two different brute-forced passwords, then the Earth Mover's Privacy with have a large value of $e^{\epsilon E_{d_\mathcal{S}}(M,M')}$ and thus, the adversary will be able to distinguish between them and thus, will know that one of those passwords are incorrect.\\\\
Although, it still doesn't mean that adversary will know which password is incorrect as there's no way of knowing the actual context, both passwords might be incorrect too, adding more obfuscation. So, we can either choose to pick one random context for the case 2, let's call it case 2.1 or we just give random decoy texts, disregarding the what context they are of, call it case 2.2. We will discuss approaches to all 3 cases in the next section.\\\\
On an abstract level, our goal is:\\
a. to classify our text with one of the categories of Brown Corpus,\\
b. to glean out the keywords from $M$ with TF-IDF,\\
c. climb up in a WordNet to get the hypernyms of the original keywords,\\
d. randomly stop in this tree-net traversal and the node we stopped at can be counted as a root, now randomly pick nouns and adjectives from the subtree of this root (those will be the hyponyms),\\
e. create a modified set (bag of words) of keywords with the nouns and adjectives picked at random, from the root hypernym of the original keywords,\\
f. run a Word2vec trained on the Brown Corpus to give embedding scores to all the elements in the modified set of keywords,\\
g. synthesize the message $M'$ with a transformer model like GPT-2, feeding it the similar words picked from the output of Word2vec,\\
NOTE: the synthesized messages $M'$s are one to one mapped with the seed value from the set $S$ of seed-space. Each encoding of $M$ uniformly at random produces a seed and corresponding to that seed, a decoy message produced from the above procedure. We can store this in form of an associative array. Decoding a seed value is just an inverse process.\\
In the end, I prove security and efficiency of this approach with the concepts of differential privacy I mentioned in this section.


\section{Method}
For the case where adversary has knowledge of the context [case 1]; to generate decoy messages, we must have a large enough repository of related words that the messages formed from those can fool the adversary.\\\\
For the case where adversary has no knowledge [case 2], we can pick a categorical corpus at random; and still the security will hold, as shown in the next section.\\\\
Therefore, our first step is to classify the message into one of the categories like ``Political'', ``Sports'', ``Financial'', ``Government Documents''. In case 2, we just randomly sample from any category for each seed encoding of the original message.
\subsection{Text Classification}
Brown corpus is partitioned into different categories. Each category contains many documents, those documents have words pertaining to the mostly used nouns, adjectives, phrases, verbs of the articles on web tagged under those categories. We first need to make a bag-of-words model for all the words of each category, and our plaintext message $M$. These bags-of-words do not contain stopwords (non-informative words) like articles, pronouns, prepositions, referential verbs; because they are almost in all the documents, in high frequency too. But contextually, they add nothing to the classification. As a part of preprocessing, we also need to remove inflection by lemmatization as tenses or suffixes might count two words as different just because of their different parts of speeches.\\\\
A naive Bayes classifier is a simple probabilistic classifier using the a posteriori decision rule in a Bayesian setting.
$$P(A|B)=\frac{P(B|A)P(A)}{P(B)}$$
Here the $A$ is category $c_i$s and $B$ is the bag-of-words of each $c_i$. Hence, the categorization of the bag-of-words for the original message [$B_M$] will be done once we train the model as follows
$$P(c_i| B_i) \propto P(B_i|c_i)P(c_i)$$
The category for $B_M$ will be computed as 
$$c_M = argmax_{c_i}P(c_i)P(B_M|c_i)$$
This is called \textit{Maximum A Posteriori} decision rule as we only used the likelihood and prior terms.
\subsection{Keyword Extraction}
The most popular approach to extract keywords from the original message $M$ is to compute each word's TF-IDF score as mentioned in previous section.
$$\text{Term Frequency of }w [TF(w)] = \frac{\#\text{word $w$ appears in } M}{\#\text{words in }M}$$
$$\text{Inverse Document Frequency of }w [IDF(w)] = \log(\frac{N}{n})$$
where $N$ is ``\#documents'' and $n$ is ``\#documents with word $w$''. Thus, it will be high for a rarer word.
$$TF-IDF(w)=TF(w)\times IDF(w)$$
Here, IDF(w) is calculated using the whole category of the corpus to which the original message $M$ is classified to (or the category which is picked uniformally at random).\\\\
The words with higher scores of $TF-IDF$ are the keywords which presents the context of $M$, not necessarily in some order or a sequence, but that will be dealt with the word embeddings. There are other techniques besides TF-IDF, like Rapid Automatic Keyword Extraction \cite{Rose:2010} which is unsupervised and thus, does not need a corpus.
\subsection{Modified Set of Keywords with WordNet}
WordNet is a lexical database of semantic relations between words. It links words into semantic relations. Those relations can be ``is-a'' type, which are called hypernyms and hyponyms. If ``dog'' is one of the keyword, then it's knowledge structure in WordNet tree consists of the hierarchy:\\
dog$\rightarrow$canine$\rightarrow$carnivore$\rightarrow$mammal$\rightarrow$$\dots$animal\\
If $U$ is a hypernym of $V$, then every $V$ is a $U$. Similarly, if $U$ is a hyponym of $V$, then every $U$ is a $V$. These synsets are arranged in a context-appropriate  hierarchies.\\
Thus, it's matter of how far we want to go up the tree and get the contextually similar words at random. We can choose to flip a coin on each level, to decide whether to go one more level up or halt on the current level, till the root of the tree is reached, in which case, the entire tree will be a candidate for the random sampling. If we halt before reaching the root, then at whatever node we stop will be the new root and that subtree will be the new tree.\\\\
With this new tree, we can again use random sampling techniques like Reservoir sampling \cite{Vitter:1985} to pick nodes.\\\\
The goal is to stay in the same context but switch the original keywords with the randomly sampled ones. While this works on heuristics, a much robust technique to modify the bag of keyword is as follows:\\\\
We have introduced Word2vec in the previous section, there are many off-the-shelf implementation available for it, which gets us the most similar words to a word given as input. The word embeddings denotes a coordinate system where related words, based on any category, are placed close together. Word2vec takes the corpus category which is chosen during text classification and the bag of keywords, it produces the modified bag of keywords where the words are replaced by the contextually similar words. As an example:\\
\textit{Input: }ice, steam\\
\textit{Output: }solid, gas, water\\
We can use the synset tree approach as a preprocessing step to the Word2vec, so that the randomness is intact. This is how we preserve the context and obfuscate the structure of the original message.
\subsection{Decoy Message Generation from Modified Set of Keywords}
\begin{figure}
  \includegraphics[width=\linewidth]{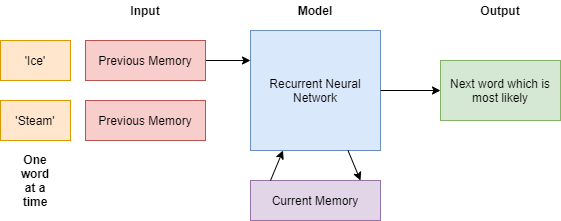}
  \caption{RNN with Long Term Dependency - LSTM}
  \label{fig:rnn}
\end{figure}

\begin{figure}
  \includegraphics[width=\linewidth]{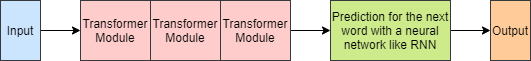}
  \caption{Text Generation with Transformer Modules and an RNN}
  \label{fig:tran}
\end{figure}

Recurrent neural networks are sometimes employed as generative models. One specific kind of RNN, called ``Long-Short Term Memory'' or LSTM takes words in an array form and from the memory of the previously presented words, text generated so far, and the current word in the queue, it can generate the next word which is the most likely. It needs to get trained on the corpus beforehand. See Figure \ref{fig:rnn}.
But LSTMs are not very polished with respect to the coherency and grammatical accuracy of the generated text. That's where the concept of transformer comes into picture. The goal of the transformer is not just to predict the next likely word, but it is trained by letting it predict ``missing words'' based on the rest of the sentence. As an example, we feed it\\
``Thomas Jefferson was an \_\_\_\_\_\_ statesman.''\\
The model is trained to fill in the blank (suppose that actual answer is ``American''), and the model is also measuring how much the words of the sentence were related to the word it predicted. The words ``was'' and ``an'' will be hardly related to ``American'' because they are used in many more contexts, but the word ``Jefferson'' is strongly related to the word ``American''.\\\\
The transformer module repeats this for every word of the corpus sentences. GPT-2 \cite{Radford:2019} has been trained on millions of web articles. It has covered every possible context already. The transformer module is encoding semantic meaning of every word, using the context around this word makes it very effective and efficient. See Figure \ref{fig:tran}, the input here is our modified bag of words.
Grover \cite{Zellers:2019} works on the same concept, except that it also always considers metadata of the words in its current memory.\\\\
If our input is,\\
\textit{``Cook the pizza in ice water for 15-20 minutes. Let it steam , then drain off the water.''}\\
After the keyword extraction, WordNet tree random sampling, word embedding and processing through transformer modules and the neural network, the result we get is:\\
\textit{1. Ice cold water is best, as the steam kills the bacteria that produces the food -borne pathogens.}\\
\textit{2. Cook ice water for 15 -20 minutes .  Add 1 tablespoon of Italian seasoning. The steam of the pressure cooker heats the seasoning.}

\section{Security}
Differential privacy in text processing is related to the topic-related contents of a message. The concept of \textit{Plausible Deniability} says that the output obtained from a query can easily have been from a database that does not consists of sender's details. That means, for the privacy of our original message $M$, producing a decoy message $M'$; it should be hard for the adversary to distinguish between the original author and our decoy message generating system. Thus, the aim here is to prove that obfuscating the origins of the context preserved message is secure.\\\\
Let $d_{\mathcal{M}}$ be a metric on $\mathcal{M}$ and $K$ a contextually similar words finding mechanism satisfying $\epsilon d_{\mathcal{M}}$-privacy:\\
$K(M)(Z)\leq e^{\epsilon E_{d_\mathcal{M}}(M,M')}K(M')(Z)$ for all $M, M' \in \mathcal{M}$ and $Z\subseteq\mathcal{M}$.\\\\
\textbf{If $K*$ is the mechanism obtained by applying $K$ independently to each element of $\mathcal{M}$, then $K*\downarrow N$ should satisfy $\epsilon NE_{d_\mathcal{M}}$-privacy for a bag-of-word of size $N$.} \cite{Fernandes:2019}\\\\
\textbf{Proof: } There are two messages $M$ and $M'$, for ease of establishing a base case, let's say both are of size $N$. We call the output of $K*$, $c$, also of size $N$. We show that $K(M)(Z)\leq e^{\epsilon E_{d_\mathcal{M}}(M,M')}K(M')(Z)$ is satisfied for a set of a singleton element $c$ and a multiplier $\epsilon N$, from there it follows that the equation is satisfied for all $Z$.\\\\
The minimization problem of the Earth Mover's Distance described in the Goal section is optimized for all values in the transportation matrix $F$ (which specifies the ``conversion'' cost of $M$ to $M'$), $F_{ij}$ are either $0$ or $1/N$.\\\\
That means, the optimal transportation for $E_{d_\mathcal{M}}(M,c)$ is achieved by comparing each word embedding score from $M$ to each word embedding score from $c$. Similarly, we also do all the comparisons for $E_{d_\mathcal{M}}(M',c)$ and thus, for $E_{d_\mathcal{M}}(M,M')$.\\\\
We introduce a new message $\underline{M}$ where each keyword in $M$ appears in $\underline{M}$. The optimal transportation for $E_{d_\mathcal{M}}(M,M')$ is\\
$E_{d_\mathcal{M}}(M,M') = \frac{1}{N}d_\mathcal{M}(\underline{M},\underline{M'})$.\\\\
Here, there is a link between $K*$ on message of size $N$ and $\underline{K}*$ on entire message space $\mathcal{M}$ by applying $K$ independently to each message. For permutation of keywords in $M$, $\sigma$, the following equality between probabilities of the contextual similarity holds:\\
$K^*(M)\{c\} = \sum_{\sigma}\underline{K}^*(\underline{M})\{\underline{c}^\sigma\}$\\
which says sum of all combinations of modified keywords will be contextually close to the original bag of keywords.
\begin{align*}
K^*(M)\{c\} &= \sum_{\sigma}\underline{K}^*(\underline{M})\{\underline{c}^\sigma\}\\
&\leq \sum_{\sigma}e^{\epsilon M_d(\underline{M},\underline{M}')}\underline{K}^*(\underline{M}')\{\underline{c}^\sigma\}\\
&= e^{\epsilon NE_d(\underline{M},\underline{M}')}\sum_{\sigma}\underline{K}^*(\underline{M}')\{\underline{c}^\sigma\}\\
&= e^{\epsilon NE_d(\underline{M},\underline{M}')}{K}^*({M}')\{{c}\}
\end{align*}
as required.

\section{Results}
\textbf{Inference Mechanism: }The adversary mechanism for author identification works by representing each decoy message $M'$ as a vector of word embeddings and comparing those to the known plaintext samples and calculating the cosine similarity with those two word embeddings. We set the threshold of that cosine score proportional to the $\epsilon$ value we choose for the differential privacy measurement.\\\\
Table \ref{tab:res} shows the results for $\epsilon$ between 0 to 30 and unmodified $\epsilon = none$ for the both messages generated by the author.
\begin{table}[]\centering
\begin{tabular}{l|ll}
$\epsilon \downarrow$\textbackslash{}\# Decoy Messages$\rightarrow$ & 100 & 500 \\ \hline
-                                                                                   & 43  & 67  \\
30                                                                                  & 34  & 46  \\
25                                                                                  & 29  & 27  \\
20                                                                                  & 16  & 13  \\
15                                                                                  & 9   & 12  \\
10                                                                                  & 2   & 4  
\end{tabular}
\caption{Amount of times the adversary can distinguish between original author's message and the decoy message for $\epsilon$-privacy}
\label{tab:res}
\end{table}
At values $\epsilon > 30$, there is not much change captured; at $\epsilon < 10$ there is not much similarity captured.
\begin{table}[]\centering
\begin{tabular}{l|ll}
$\epsilon \downarrow$\textbackslash{}\# Decoy Messages$\rightarrow$ & 100 & 500 \\ \hline
-                                                                                   & 23  & 43  \\
30                                                                                  & 19  & 46  \\
25                                                                                  & 22  & 41  \\
20                                                                                  & 20  & 36  \\
15                                                                                  & 17   & 38  \\
10                                                                                  & 16   & 33  
\end{tabular}
\caption{Amount of times the adversary can distinguish between original message's context and the decoy message's context for $\epsilon$-privacy}
\label{tab:res1}
\end{table}
Table \ref{tab:res1} shows more consistency with a distinguisher which is distinguishing the contexts of the messages with comparison to a distinguisher which is distinguishing between source of the messages.\\\\
That can be because the discriminator neural network used for the context distinguishing had more data to work on (entire corpus) as opposed to smaller amount of user's data which was available to us (a particular news website also varies widely in style of writing because there are multiple authors, this is the concept Grover works on).

\section{Conclusion and Future Work}
I showed how natural language processing can be used to generate contextually similar decoy messages in Honey Encryption scheme, with author identity and message structure obfuscation. With large enough available data on the context similar to the message, we can identify keywords of that message, to get similar words having closer word embeddings and then feed it to an already trained transformer-neural network combination, to generate realistic decoy messages.\\\\
I proved it used $\epsilon$-differential privacy mechanisms. I also showed the trade-off between author discrimination and context discrimination.\\\\
Naturally, with supervised learning, we need more and more data to achieve higher levels of obfuscation. GPT-2 is trained on millions of webpages, but author styles will vary widely, if not the context of the messages.\\\\
Future work can be focused towards extracting the context from an unsupervised learning technique, so that the requirement of large corpus data on a subject which is not much discussed on, can be eliminated. That will mean less processing time as well.\\\\
We can also find a method to bring down the author identification probability, in case adversary has lots of publicly data available for a certain author.




\end{document}